\documentclass{iaus}
\usepackage{graphicx}

\title[X-ray spectroscopy of BD +30$^\circ$3639] 
{High-resolution X-ray Spectroscopy \break of BD +30$^\circ$3639}

\author[Kastner et al.]   
{Joel H. Kastner$^1$, Young Sam Yu$^1$, John Houck$^2$, Ehud
  Behar$^3$, \break Raanan Nordon$^3$, %
\and Noam Soker$^3$}

\affiliation{$^1$Center
  for Imaging Science, Rochester Institute of Technology,
  Rochester, NY 14623, USA; email: 
  jhk@cis.rit.edu\\[\affilskip] 
$^2$MIT Kavli Institute, Massachusetts Institute of
  Technology, Cambridge, MA 02139, USA  \\[\affilskip]
$^3$Department of Physics, Technion - Israel Institute of
  Technology, Haifa 32000, Israel} 

\pubyear{2006??}
\volume{234}  
\date{?? and in revised form ??}
\editors{R. Mendez et al., eds.}
\begin{document}

\maketitle

\begin{abstract}
We present preliminary results from the first X-ray gratings
spectrometer observations of a planetary nebula (PN). We have used the
Chandra X-ray Observatory Low Energy Transmission Gratings
Spectrometer (LETGS) to observe the bright, diffuse X-ray source
within the well-studied BD +30$^\circ$3639. The LETGS spectrum of BD
+30$^\circ$3639 displays prominent and well-resolved emission lines of
H-like C, O, and Ne and He-like O and Ne. Initial modeling indicates a plasma
temperature $T_X \sim 2.5\times10^6$ K and abundance ratios of C/O
$\sim20$, N/O $\stackrel{<}{\sim}1$, Ne/O $\sim4$, and Fe/O
$\stackrel{<}{\sim}0.1$. These results suggest that the X-ray-emitting
plasma is dominated by the shocked fast wind from the emerging PN
core, where this wind gas likely originated from the intershell region
of the progenitor asymptotic giant branch star.  
\end{abstract}

\firstsection 

\section{Introduction}

The Chandra X-ray Observatory (CXO) and XMM-Newton X-ray Observatory
have ushered in a new era in the study of X-ray emission from
planetary nebulae (see review by Guerrero in these proceedings),
providing new insight into wind-wind interactions in PNs.  The CXO and
XMM discoveries of diffuse X-ray emission within PNs such as BD
+30$^\circ$3639 (Kastner et al.\ 2000), NGC 6543 (Chu et al.\ 2001),
NGC 7009 (Guerrero et al.\ 2002), and NGC 40 (Montez et al.\ 2005) are
indicative of the interaction of a quasi-spherical fast wind from the
newly unveiled central star with the former asymptotic giant branch
(AGB) wind, whereas the X-ray morphologies of NGC 7027 (Kastner et
al.\ 2001) and Menzel 3 (Kastner et al.\ 2003) are indicative of the
presence and shaping action of collimated outflows. 

The source of the X-ray emitting gas in PNs remains to be
determined, however. It is thus intriguing that the X-ray
data provide indications of significant abundance anomalies
in the superheated plasma within PNs. In particular, greatly enhanced
abundances of O and Ne and a large depletion of Fe (relative
to solar) are suggested by CCD X-ray spectroscopy of BD
$+30^\circ$3639 (Arnaud et al. 1996; Kastner et al.\ 2000;
Maness et al.\ 2003). These results stand in sharp contrast
to observations at optical and infrared wavelengths, which
show depleted Ne in the optically bright shell of this PN.
Indications of abundance anomalies are also observed in NGC
6543 (Chu et al.\ 2001) and NGC 7027 (Maness et al.\ 2003). 
On the other hand, Georgiev et al.\ (2006) contend
that the extant X-ray CCD spectra cannot
provide useful constraints on the abundances of
X-ray-emitting plasmas within PNs. Indeed, the various
analyses cited above have arrived at very different results
concerning plasma adundances within the hot bubble of BD
$+30^\circ$3639, indicative of the large
uncertainties inherent in spectral modeling that relies on relatively
low resolution X-ray CCD spectra.

To make progress on these and other problems that have
surfaced as a consequence of the recent X-ray detections of PNs
by CXO and XMM, we require observations combining high spatial
and spectral resolution, so as to infer gas temperature and
composition as a function of position within the X-ray
emitting plasma.  Just prior to this conference, we obtained
such an observation of the well-studied BD $+30^\circ$3639
-- the brightest diffuse X-ray source among PNs -- using
Chandra's Low Energy Transmission Gratings spectrometer in
combination with its Advanced CCD Imaging Spectrometer
(LETG/ACIS-S). Here, we report on preliminary results
obtained from this, the first X-ray gratings spectrometer
observation of a planetary nebula.

\section{Observations and Results}

We obtained the first of two 150 ks LETG/ACIS-S
observations\footnote{The second 150 ks observation is
presently scheduled for 2007 January.} of BD +30$^\circ$3639 in 2006 February
(86 ks) and March (61 ks). 
We used standard Chandra X-ray Center spectral calibration
and analysis tools\footnote{http://cxc.harvard.edu/ciao/} to update
calibrations and to extract and
merge the first-order LETG spectra. 
Defaults were used for all
spectral extraction parameters (e.g., cross-dispersion region width and
first-order event pulse height ranges) except for the
spectral bins, which were rebin to 0.126 \AA.
Corresponding LETG/ACIS-S spectral sensitivity and spectral resolution
calibration files were constructed in parallel with the
spectral extraction. 

\begin{figure}
 \includegraphics[scale=0.35]{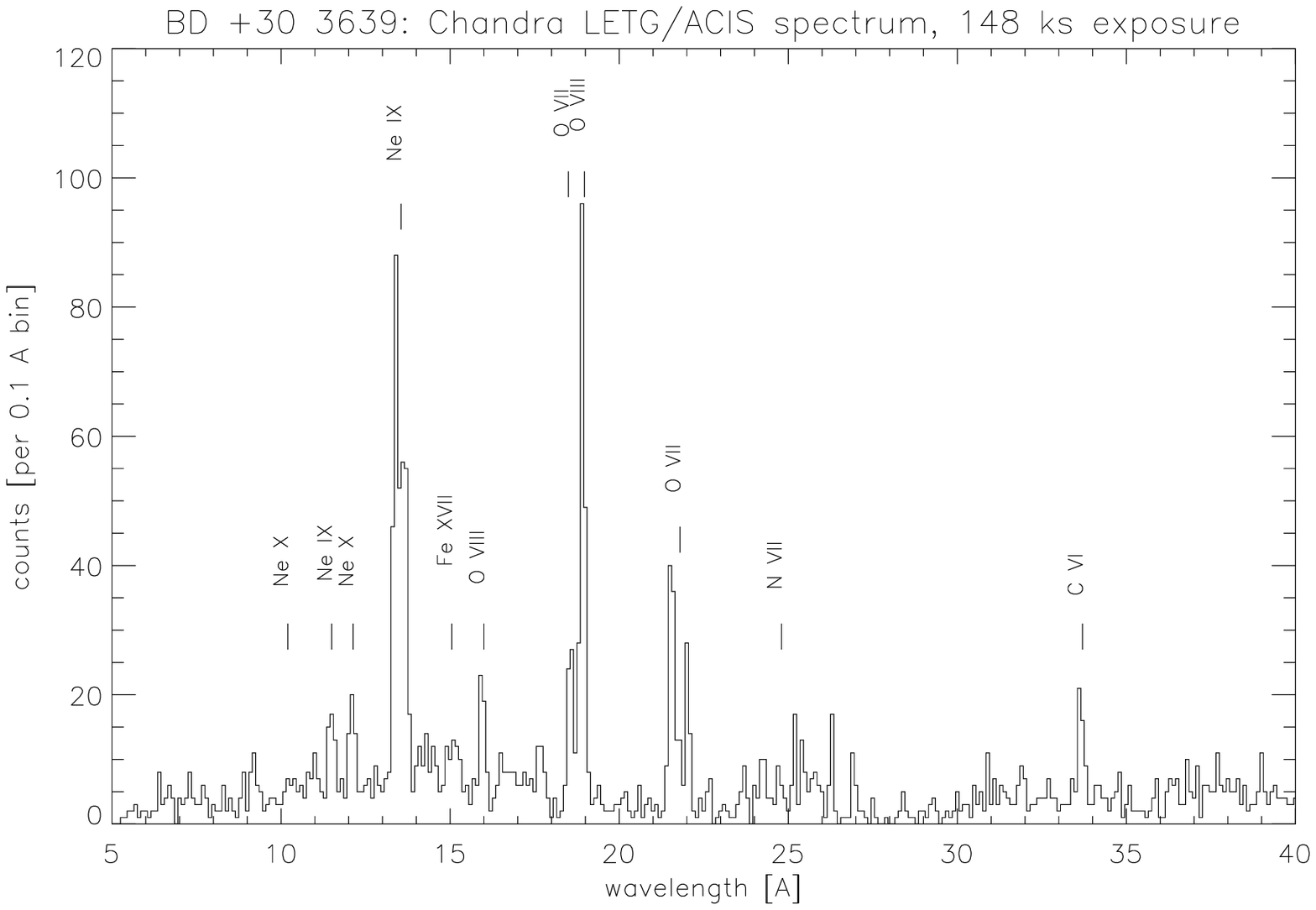}
 \includegraphics[scale=0.35]{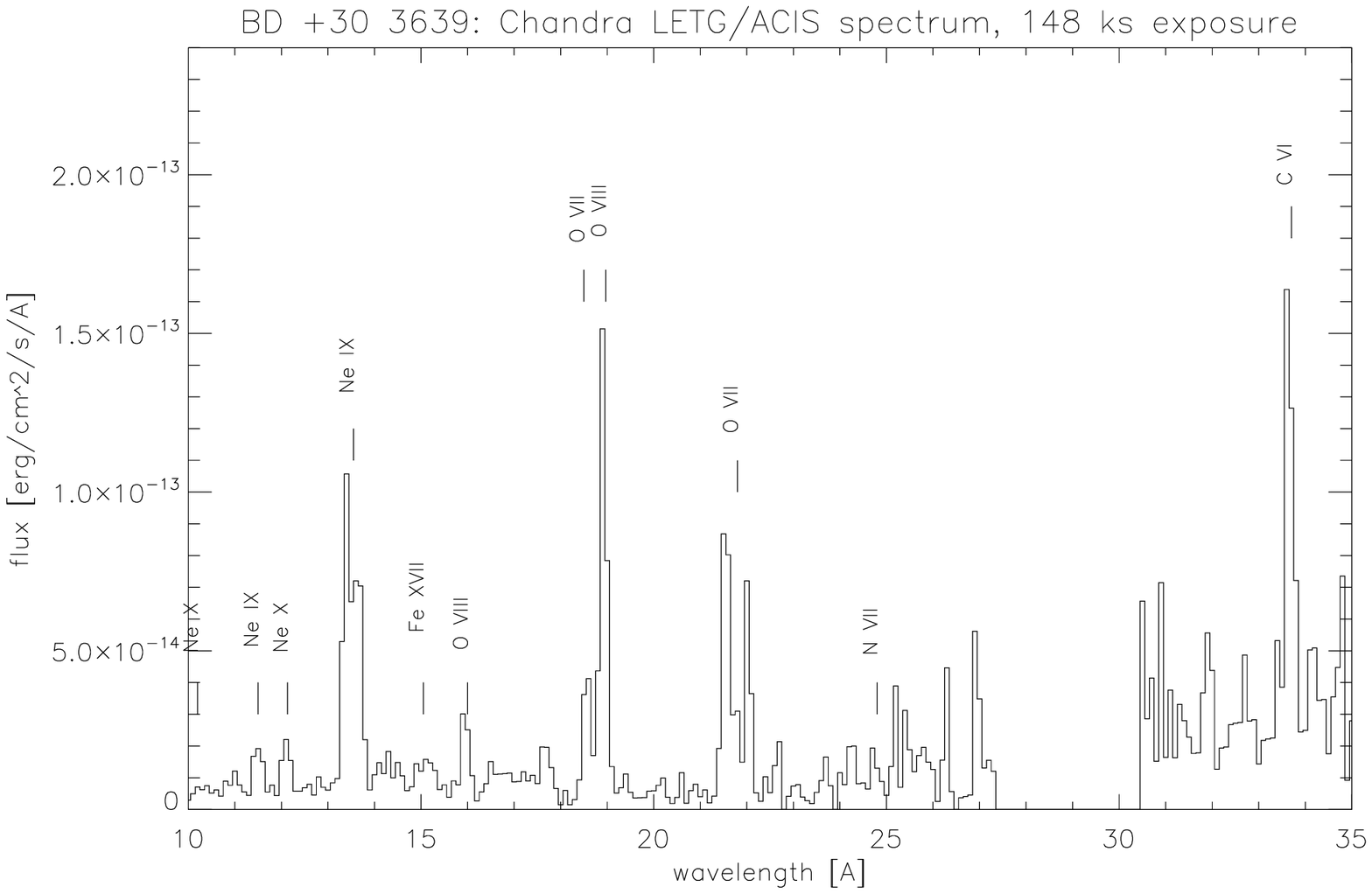}
  \caption{{\it a)} Combined positive and negative first-order
  LETG/ACIS-S counts spectrum of BD +30$^\circ$3639. {\it
  b)} Flux-calibrated first-order LETG/ACIS-S spectrum, in
  energy units. The apparent rising ``continuum'' at $\lambda > 25$
  \AA\ is likely the result of low photon detection efficiency combined with
  incomplete background subtraction.}  
\end{figure}

Fig.\ 1a shows the resulting, first-order
LETG/ACIS-S counts spectrum of 
BD +30$^\circ$3639. The brightest lines in the spectrum, in
terms of total counts, are the resonance line of O
{\sc viii} ($\lambda$ 18.97), the He-like triplet line
complex of Ne {\sc ix} ($\lambda\lambda$ 13.45, 13.55,
13.7), and the He-like triplet of O {\sc vii}
($\lambda\lambda$ 21.60, 21.80, 22.10). The 
resonance line of C {\sc vi} ($\lambda$ 33.6) is also
detected, as are various other, weaker lines of H-like and
He-like O and Ne. Lines of H-like N (e.g., the 
resonance line at $\lambda$ 24.78) and highly ionized Fe are
notably weak or absent. The flux-calibrated spectrum,
obtained as the efficiency-weighted average of the positive
and negative first-order spectra after correction for their
respective efficiencies, is displayed in Fig.\ 1b. The
strength of C {\sc vi} relative to the strongest lines of Ne
{\sc ix}, O {\sc vii}, and O {\sc viii} is apparent in this
flux-calibrated spectrum.

\begin{figure}
\centering{\includegraphics[scale=0.42]{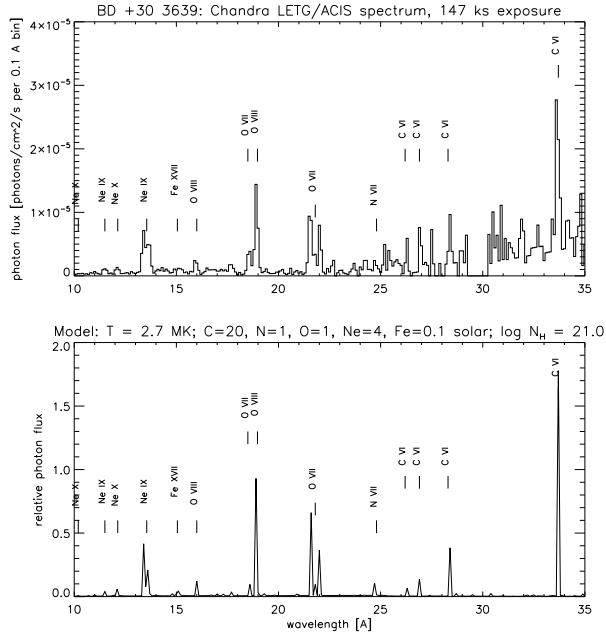}}
  \caption{Observed LETG/ACIS-S spectrum of BD
  +30$^\circ$3639 in photon flux units {\it (top panel)} and 
  ISIS model {\it (bottom panel)}. }
\end{figure}

Theoretical ratios for the resonance lines of H-like to
He-like species of O and Ne obtained from the Astrophysical
Plasma Emission Database (APED\footnote{http://cxc.harvard.edu/atomdb/};
Smith et al.\ 2001) as well as the ``G ratio'' calculated
from the triplet complex of He-like O {\sc vii} (Smith et
al.) indicate a plasma temperature in the relatively narrow
range $T_x = (2.4-2.8)\times10^{6}$ K. In light of these
results for $T_x$, 
we used the Interactive Spectral Interpretation System
(ISIS\footnote{http://space.mit.edu/CXC/ISIS/}) to construct
a series of APED models of varying plasma elemental
abundances with $T_x$ fixed at $2.7\times10^{6}$ K and the
intervening absorbing column fixed at 
$\log{N_H} ({\rm cm^{-2}}) = 21.0$ (the latter constraint is
based on results from 
ACIS-S X-ray CCD spectroscopy, and is consistent with
measurements of visible-wavelength extinction; Arnaud et
al.\ 1996, Kastner et al.\ 2000). With
$T_x$ and $N_H$ so constrained, we
find good agreement between data 
and model for abundance ratios (relative to solar) of C/O =
20, Ne/O = 4, N/O = 1, and Fe/O = 0.1 (Fig.\ 2). The last
two abundance ratios should be interpreted as upper limits,
as the N and Fe lines are not well detected. Although
these results (and, in particular, the inferred C/O ratio)
are somewhat dependent on the assumed value of $N_H$, we estimate a
relatively firm upper limit of C/O $\sim40$, corresponding
to $\log{N_H} ({\rm cm^{-2}}) = 21.4$ (the largest value of
$N_H$ obtained from X-ray CCD spectroscopy of BD
+30$^\circ$3639; Maness et al.\ 2003). We also cannot yet rule out
the presence of a lower-$T_x$ component in the plasma, in
which case the C/O abundance ratio would still be enhanced
but is very unlikely to be $>20$ with respect to solar.  

\section{Discussion}

The sharply non-solar composition of the shocked,
X-ray-emitting plasma in BD +30$^\circ$3639 strongly
suggests this gas originated in an AGB star ``intershell''
region (Herwig 2005 and references
therein). Specifically, the measured C overabundance
corresponds to a C/O number ratio of $\sim10$ (as compared
with C/O $\sim1.6$ in the nebular gas; Pwa et al.\ 1986).
This C/O number ratio is consistent with the predictions of
models that describe He shell burning and subsequent
dredge-up into the layer between the He- and H-burning
shells of the former AGB star. In addition, the combination
of strong Fe depletion and Ne enhancement can be readily
explained as due to the $s$ process within the
``pulse-driven convection zone'' (PDCZ) associated with the
He-burning shell within the former AGB star. 
Formation of excess Ne would occur due to alpha capture on $^{14}$N
and then $^{18}$O. The resulting $^{22}$Ne then serves as a
neutron source for the $s$ process within the PDCZ, thereby
depleting Fe.

The temperature of shocked gas probed by lines of H- and
He-like O and Ne is $T_x\sim2.5\times10^6$, confirming an
earlier estimate of $T_x$ we obtained by modeling the
lower-resolution X-ray CCD spectrum of BD +30$^\circ$3639
(Kastner et al.\ 2000). This plasma temperature is lower
than expected from a simple adiabatic shock model, given the
present central star wind speed of $v_{f}\sim700$ km
s$^{-1}$. Among the numerous potential explanations for this
discrepancy (see Soker \& Kastner 2003 and references therein),
mixing of nebular gas with the shocked fast wind 
appears to be ruled out as a plasma ``coolant,'' given the
highly non-nebular abundance signatures apparent in the
LETG/ACIS-S spectrum. On the other hand, these observations cannot
rule out heat conduction as a mechanism for moderating
$T_x$; further numerical calculations like those
conducted by Steffen et al.\ (these proceedings)
should examine this mechanism, in light of our results. It
also remains possible that the shocked wind presently seen
radiating in X-rays was ejected at an earlier epoch, when
the fast wind velocity was lower ($v_{f}\sim450$ km
s$^{-1}$); such an explanation, initially proposed by Arnaud
et al.\ (1996), has recently been elaborated on by Soker \&
Kastner (2003) and Akashi et al.\ (2006).




\begin{acknowledgments}
This research was supported by NASA
via Chandra Award
GO5--6008X issued to R.I.T. by the CXO Center,
which is operated on behalf of NASA by Smithsonian
Astrophysical Observatory under contract NAS8--03060.
\end{acknowledgments}

\end{document}